\documentstyle[12pt]{article}
\begin{document}
\pagestyle{plain}

\newcommand{\be}{\begin{equation}}
\newcommand{\ee}{\end{equation}}
\newcommand{\bea}{\begin{eqnarray}}
\newcommand{\eea}{\end{eqnarray}}
\newcommand{\vp}{\varphi}
\newcommand{\pr}{\prime}
\newcommand{\sech} {{\rm sech}}
\newcommand{\cosech} {{\rm cosech}}
\newcommand{\psib} {\bar{\psi}}
\newcommand{\cosec} {{\rm cosec}}

\title{{Stationary Solitons of the Fifth Order KdV-type}\\
{Equations and Their Stabilization}}
\author{
B. Dey \\
{\small \sl Department of Physics, University of Poona,}\\
{\small \sl Ganeshkhind, Pune 411 007, India}\\
{\small \sl }\\
\and Avinash Khare \\
{\small \sl Institute of Physics, Sachivalaya Marg,}\\
{\small \sl Bhubaneswar - 751005, India}\\
{\small \sl  }\\
\and C. Nagaraja Kumar \\
{\small \sl School of Physics, University of Hyderabad,}\\
{\small \sl Hyderabad - 500 046, India}}
\maketitle

\begin{abstract}

Exact stationary soliton solutions of the fifth order KdV type equation
$$ u_t +\alpha u^p u_x +\beta u_{3x}+\gamma u_{5x} = 0$$
are obtained for any p ($>0$) in case $\alpha\beta>0$, $D\beta>0$, $\beta\gamma<0$ 
(where D is the soliton velocity), and it is shown that these solutions are 
unstable with respect to small perturbations in case $p\geq 5$. Various 
properties of these solutions are discussed. In particular, it is shown that 
for any p, these solitons are lower and narrower than the corresponding 
$\gamma = 0$ solitons. Finally, for p = 2 we obtain an exact stationary 
soliton solution even when $D,\alpha,\beta,\gamma$ are all $>0$ and discuss
its various properties.
\end{abstract}
\pagebreak

In recent years the soliton solutions of the fifth order KdV-type highly
nonlinear equations 
\be\label{1}
{\partial u\over\partial t} + \alpha u^p {\partial u\over\partial x} + \beta
{\partial^3 u\over\partial x^3} +\gamma {\partial^5 u\over\partial x^5} = 0
\ee
(with p $>0$) have received considerable attention in the literature [1]. In 
particular, attention has been focused on the role of the last term in this 
equation which describes higher order dispersive effects
and may have important influence on the properties of the solitons. The 
equation with p = 1 is the fifth order KdV eqution [2] which has applications 
in fluid mechanics (e.g., shallow water waves with surface tension ), plasma 
physics etc. On the other hand, for p = 2 one has fifth order MKdV equation 
which may also be of interest in fluid mechanics, plasma
physics. Finally, above equation with $p\geq 4$ has considerable theoretical 
interest in connection with the general problem of collapse of nonlinear 
waves. Indeed, it is well known
that the stationary soliton solutions of the eq.(\ref{1}) for $\gamma = 0$  
and $\alpha > 0, \beta > 0$ as given by 
\be\label{2}
u_0 (\xi) = \bigg ( {D\over 2} (p+1)(p+2)\bigg )^{1/p} 
\sech^{2/p} ({p\over 2}\sqrt D\xi)
\ee
are unstable with respect to collapse-type instabilities if 
$p\geq 4$ [3]. Here $\xi = x - Dt$ 
with the velocity $D > 0$ and without any loss of generality we have chosen 
$\alpha = \beta =1$ throughout
this paper unless mentioned otherwise. The role of the last term in 
eq.(\ref{1}) has been discussed by several people including Karpman [1]. Based 
on analytical and numerical
work, it has been suggested that the fifth order term stabilizes  the soliton
specially for $p \leq 6$ [4]. It has also been conjectured that for large 
enough $\mid\gamma\mid$ even soliton solutions with $p > 6$ could also be 
stable. Unfortunately, no analytical soliton solution is known in literature 
when $\gamma\not = 0$ except when p = 1 [5] and hence so far it
has not been possible to check the validity of these conjectures.

Recently Hai and Xiao [6] have obtained a soliton solution of eq.(\ref{1}) 
with p = 1 which is valid to first order in 
$\mid\gamma\mid$ and have shown that the corresponding soliton is lower and 
narrower then the unperturbed ($\gamma$ = 0) soliton as given by eq.(\ref{2}). 
Is their conclusion
also valid for large $\mid\gamma\mid$ ? Further, is it also true for any p?

It is thus clearly of great interest to obtain an exact analytical stationary 
soliton solution of eq.(\ref{1}) and test the validity of these conjectures.
The purpose of this note is to show that an exact solution of eq.(\ref{1}) is
\be\label{3}
u(\xi) = \bigg ({D\over 8} {(p+1)(p+4)(3p+4)\over (p+2)}\bigg )^{1/p} 
\sech^{4/p} \bigg ({p\xi\sqrt{D(p^2+4p+8)}\over 4 (p+2)}\bigg )
\ee
where $\xi = x-Dt$, $\alpha = \beta = 1, D > 0, \ \gamma < 0$ and 
\be\label{3a}
\epsilon \equiv (D\mid\gamma\mid)^{1/2} = {2(p+2)\over (p^2+4p+8)} < 1
\ee
Several properties of these solutions are discussed and it is shown that 
contrary to the expectation
based on the numerical studies [4], these soliton solutions continue to 
be unstable in case $p \geq 5$. We also show that for any p, the soliton 
as given by eq.(\ref{3}) is indeed
lower and narrower than the unperturbed soliton as given by (\ref{2}) 
no matter what the value of $\mid\gamma\mid$ is.

Finally we also present an another stationary soliton solution of 
eq.(\ref{1}) in case p = 2 for which $\alpha,\beta,\gamma, D > 0$. 
This is interesting because recently
there have been suggestions in the literature that a stationary soliton 
solution to eq.(\ref{1}) may not exist in case $\alpha,\beta,\gamma, 
D > 0$ [7]. 

Let us consider eq.(\ref{1}) and look for stationary soliton solutions 
of the form u = u $(\xi)$
where $\xi = x - Dt$ with the boundary condition that $u\rightarrow 0$ as 
$\xi\rightarrow \pm\infty$. 
On integrating eq.(\ref{1}) with respect to $\xi$ and choosing the 
constant of integration to be zero we have
\be\label{5}
-Du +{u^{p+1}\over (p+1)} +{d^2u\over d\xi^2} +\gamma {d^4 u\over d\xi^4} 
= 0\
\ee
We now look for a solution of the form
\be\label {6}
u (\xi) = A \sech^b (m \xi)
\ee
On using eq.(\ref{6}) in eq.(\ref{5}) it is easily shown that $b = 4/p, D > 0, 
\ \gamma < 0$ and as given by eq.(\ref{3a}) and 
\be\label{6a}
A = \bigg ( {D(p+1)(p+4)(3p+4)\over 8(p+2)}\bigg )^{1/p}, 
\ m = {p\sqrt{D(p^2+4p+8)}\over 4(p+2)}
\ee
so that the solution is as given by eq.(\ref{3}). Some of the interesting 
properties of this solution are
 
(i) the solution is fairly localized and is truly a nonperturbative solution
in the sense that as $\mid\gamma\mid \longrightarrow 0$ the solution diverges 
rather than
tending to the third order KdV soliton solution as given by eq.(\ref{2}).

(ii) for this solution the wave travels only to the left $(D > 0)$ with 
amplitude $\propto (velocity)^{1/p}$ and hence taller the wave, faster it 
moves! Also notice 
from eq.(\ref{3a}) that the velocity $\propto 1/\mid\gamma\mid$ and hence 
smaller the $\mid\gamma\mid$, larger is the velocity and vice a versa.
 
(iii) From the solution (\ref{3}) we observe that
\be\label{7}
\int^{\infty}_{-\infty} u^{p/2} (x) dx = {2\sqrt 2\over p}
\bigg ( {(p+1)(p+2)(p+4)(3p+4)\over (p^2+4p+8)}\bigg )^{1/2} = constant.
\ee
which is independent of D and hence $\mid\gamma\mid$. Thus so far as the 
dependence on $\mid\gamma\mid$
(or D) is concerned, one can say that $u(x)\sim [\delta (x)]^{2/p}$.

(iv) Note that solution (\ref{3}) with $\xi = x - Dt$ replaced by 
$x - Dt+x_0$ with
$x_0$ being an arbitrary constant is also a solution to eq.(\ref{1}). 

(v) On comparing the two soliton solutions as given by eqs.(\ref{2}) and 
(\ref{3}) we find that irrespective of the value of $\mid\gamma\mid$ 
( and hence D), the $\mid\gamma\mid \not = 0$ soliton
is lower than the $\gamma$ = 0 soliton. For example, the amplitude difference 
of the two is given by
\be\label{7a}
(amp)_{\gamma = 0} - (amp)_{\mid\gamma\mid} = \bigg ({D\over 2} 
(p+1)(p+2)\bigg )^{1/p} \bigg (1-({(p+4)(3p+4)\over 4(p+2)^2})^{1/p}\bigg ) > 0
\ee
which is equal to ${D\over 12}$ for p = 1 and increases with p. Similarly, on 
comparing the two solutions (\ref{2}) and (\ref{3}) it is easily seen that the
$\gamma \neq 0$ soliton is narrower than the $\gamma = 0$ soliton i.e. 
$u < u_0$ for any $\xi \approx 0$.
 
(vi) As remarked by Karpman [1], eq.(\ref{1}) is a Hamiltonian system for 
which the energy and momentum are given by
\be\label{8}
E_s = \int^{\infty}_{-\infty} dx \bigg ( {1\over 2} ({du\over dx})^2 - 
{u^{p+2}\over (p+1)(p+2)} - {\gamma\over 2}({d^2u\over dx^2})^2 \bigg )
\ee
\be\label{9}
P_s ={1\over 2} \int^{\infty}_{-\infty} dx u^2
\ee
For the solution (\ref{3}) we find that
\be\label{10}
E_s = {D A^2 2^{8/p} \Gamma^2(4/p)\over 4m \Gamma (8/p) (p+2)^2(p+4)} 
\bigg ((p+2)^2(p-4) - {8p^3 (p+3)\over (p+8)(3p+8)}\bigg )
\ee
\be\label{11}
P_s = {A^2 2^{8/p} \Gamma^2(4/p)\over 4m \Gamma(8/p)}
\ee
where A and m are as given by eq.(\ref{6a}). Thus we find that $E_s \propto 
D^{({2\over p}+{1\over 2})}$ while $P_s \propto D^{({2\over p}-{1\over 2})}$ 
and further $E_s < 0$ for $p\leq 4$ while $E_s > 0$ for $p\geq 5$
while $P_s > 0$ for any $ p(>0)$.

Let us now address the question of the stability of the soliton solution 
(\ref{3}) with respect to small perturbation. Based on a conjecture which 
is supported by some numerical results, Karpman obtained the following 
sufficient condition for the soliton stability in case $\gamma < 0$ [1] 
\be\label{12}
({\partial P_s\over\partial D})_{\gamma} > 0
\ee
Based on some numerical work, Karpman then conjectured [1] the stability of 
the $\gamma < 0$ soliton solutions for at least $p\leq 6$ in case 
$0 < \epsilon < \epsilon_p$ depending on p. Does our solution support 
Karpman's conjecture ? On using the fact that for our solutions
$P_s \propto D^{({2\over p}-{1\over 2})}$ it follows that 
\be\label{13}
({\partial P_s\over\partial D})_{\gamma} 
= {1\over D} ({2\over p}-{1\over 2}) P_s
\ee
so that $({\partial P_s\over\partial D})_{\gamma} > 0$ if and only if $p < 4$. 

Karpman has also given a simple and useful necessary condition for the soliton 
stability given by [1]
\be\label{14}
R \equiv  {\mid\gamma\mid J_2\over 2D P_s} > 
{p(p-4)\over (p^2+4p+32)} \equiv R_{cr} (p)
\ee
where
\be\label{15} 
J_2 = \int^{\infty}_{-\infty} dx ({\partial^2 u\over\partial x^2})^2
\ee
For the solution (\ref{3}), we find that 
\be\label{16} 
J_2 = {128 A^2 m^3 (p+3) 2^{8/p}\Gamma^2 (4/p)\over p^{2}(3p+8)(p+8)\Gamma(8/p)}
\ee
where A and m are as given by eq.(\ref{6a}). Using eqs.(\ref{11}) and 
(\ref{3a}) we then 
find that the fifth order stationary soliton solution is stable provided
\be\label{17} 
{4p^2(p+3)\over (p+2)^2(p+8)(3p+8)} > {p(p-4)\over (p^2+4p+32)}
\ee
i.e if $3p^5+28p^4-608p^2-1664p - 1024 < 0$. We find from here that the 
$\gamma\not = 0$ soliton is stable so long as $p \leq 4.75$. Since 
$\gamma$ = 0 soliton was only stable for $p\geq 4$ hence
it is clear that the fifth order term has increased the stability range 
but not by as much as it had been conjectured. In particular, for $p\geq 5$ 
the soliton solution (\ref{3}) is still unstable under small perturbations. 

Finally, for the special case of p = 2, we display another 
stationary soliton solution. In particular, on using the ansatz 
u = A $\sech (m\xi)\tanh (m\xi)$ in eq.(\ref{1})
it is easily shown that an exact stationary soliton solution is
\be\label{18}
u(\xi) = ({360D\over 11})^{1/2} \sech (\sqrt{{100\over 11}}\xi) 
\tanh (\sqrt{{100\over 11}}\xi)
\ee
provided $\gamma, D > 0,$ ($\alpha = \beta = 1$ as usual ) and 
$\xi \equiv (D\gamma)^{1/2}$ =$\sqrt{{11\over 10}}$.
Note that for the p = 2 soliton of the type (\ref{3}), $\epsilon = 2/5$. The 
solution (\ref{18}) is again a truly localized nonperturbative solution for 
which amplitude $\propto (velocity)^{1/2}$. Note that u not only vanishes as 
$\xi\rightarrow \pm\infty$ but also at $\xi = 0$ and further $u(\xi)$ 
is negative for $\xi < 0$. For this solution $P_s = 12\sqrt{{100\over 11}}$ 
while $E_s= -{60\over 77}\sqrt{{10 D^3\over 11}}$. Further, it is easily seen 
that this soliton is also lower
and narrower than the corresponding $\gamma = 0$ (and p = 2) soliton as given 
by eq.(\ref{2}).

Before ending this note we would like to exhibit a stationary kink solution to 
eq.(\ref{1}) in case
p = 4, $D,\alpha < 0, \beta, \gamma > 0$ (or $D, \alpha >0 $ and 
$\beta,\gamma < 0$) and $\epsilon \equiv (\mid D\mid\gamma)^{1/2}$ 
= $\sqrt 6/10$. The solution is 
\be\label{19}
u(\xi) = (\mid D\mid )^{1/4} \tanh (\sqrt{{5\mid D\mid\over 6}}\xi)
\ee
where without any loss of generality, we have chosen $\alpha = -1,\beta = 1$.

\noindent {\bf Acknowledgments}

One of us (CNK) thanks Prof. J. Hietarinta and Prof. J. Satsuma for useful 
discussions during CIMPA School in Jan.96 in India. CNK's work is supported 
by CSIR through R. A. scheme.
\vfill
\eject


\begin{thebibliography}{99}
\bibitem{[1]} V.I. Karpman, Phys. Lett. {\bf A210} (1996) 77 and references 
therein.
\bibitem{[2]} For the p =  1 case see T. Kawahara, J. Phys. Soc. Jpn. {\bf 33} 
(1972) 360; Y. Pomeau, A. Ramani and B. Grammaticos, Physics {\bf D31} (1988) 
127; J.K. Hunter and J.-M. Vanden-Broeck, J. Fluid
Mech. {\bf 134} (1983) 205, K.A. Gorshkov and L.A. Ostrovsky, Physica {\bf D3} 
(1981) 424; A. Grimshaw and B.A. Malomad, J. Phys. {\bf A26} (1993), 4087.
\bibitem{[3]} E.A. Kuznetsov, Phys. Lett. {\bf A101} (1984) 314;
E.A. Kuznetsov, A.M. Rubenchik and V.E. Zakharov, Phys. Rep. {\bf 142} (1986) 
103; V.E. Zakharov, ed. in: Proc. Int. Workshop on wave collapse Physics, 
Physica {\bf D52} (1991) 1.
\bibitem{[4]} V.I. Karpman, ref.1, V.I. Karpman and J.-M Vanden-Broeck, Phys. 
Lett. {\bf A200} (1995) 423.
\bibitem{[5]}  K. Nozaki, J. Phys. Soc. Jpn. {\bf 56} (1987) 3052.
\bibitem{[6]} W. Hai and Y. Xiao, Phys. Lett. {\bf A208} (1995) 79.
\bibitem{[7]} V.I. Karpman, Phys. Lett. {\bf A186} (1994) 300, 303.
\end{thebibliography}
\end{document}